\documentclass[letter,		  %
               keeplastbox,   
               ]{jacow}
\usepackage{pdfpages,multirow,ragged2e} %
\makeatletter%
	\ifboolexpr{bool{xetex}}
	 {\renewcommand{\Gin@extensions}{.pdf,%
	                    .png,.jpg,.bmp,.pict,.tif,.psd,.mac,.sga,.tga,.gif,%
	                    .eps,.ps,%
	                    }}{}
\makeatother

\ifboolexpr{bool{xetex} or bool{luatex}} 
 {}                                      
 {\usepackage[utf8]{inputenc}}           

\usepackage[USenglish]{babel}

\ifboolexpr{bool{jacowbiblatex}}%
 {%
  \addbibresource{jacow-test.bib}
  \addbibresource{biblatex-examples.bib}
 }{}
\begin{document}

\title{Phase Space Density Evolution in MICE}

\author{D. Rajaram\thanks{durga@fnal.gov}, Illinois Institute of Technology, Chicago, USA \\
		V. Blackmore, Imperial College London, London, UK \\
		\it{on behalf of the MICE collaboration}}
\maketitle

\begin{abstract}
 The Muon Ionization Cooling Experiment (MICE)  will demonstrate the feasibility of ionization cooling, the technique by which it is proposed to cool the muon beam at a future neutrino factory or muon collider. The position and momentum reconstruction of individual muons in the MICE trackers allows for the development of alternative figures of merit in addition to beam emittance. Contraction of the phase space volume occupied by a fraction of the sample, or equivalently the increase in phase space density at its core, is an unequivocal cooling signature. Single-particle amplitude and non-parametric statistics provide reliable methods to estimate the phase space density function. These techniques are robust to transmission losses and non-linearities, making them optimally suited to perform a quantitative cooling measurement in MICE.
\end{abstract}

\section{INTRODUCTION}

Muon beams of low emittance have been proposed as the basis for  intense, well-characterized neutrino beams for the Neutrino Factory and for lepton-antilepton collisions at energies of up to several TeV at the Muon Collider~\cite{ref:nf, ref:nf2, ref:nfmc}. The stored muons at such facilities originate from decays of pions and therefore inherit a large volume in phase space. For efficient acceleration the phase space volume (emittance) must be reduced (cooled) significantly to fit within the acceptance of a storage ring or accelerator beam pipe. Due to the short lifetime of the muon, ionization cooling is the only practical technique by which to cool beams of muons~\cite{ref:neuffer, ref:cooling-methods}.

In ionization cooling, a muon beam passes through an absorber material losing momentum in both transverse and longitudinal dimensions, thereby reducing the RMS emittance and increasing its phase space density. Subsequent acceleration though rf cavities restores the longitudinal momentum, thus resulting in a net reduction of the transverse emittance. 
\begin{figure*}[!htb]
   \centering
    \includegraphics*[width=\textwidth]{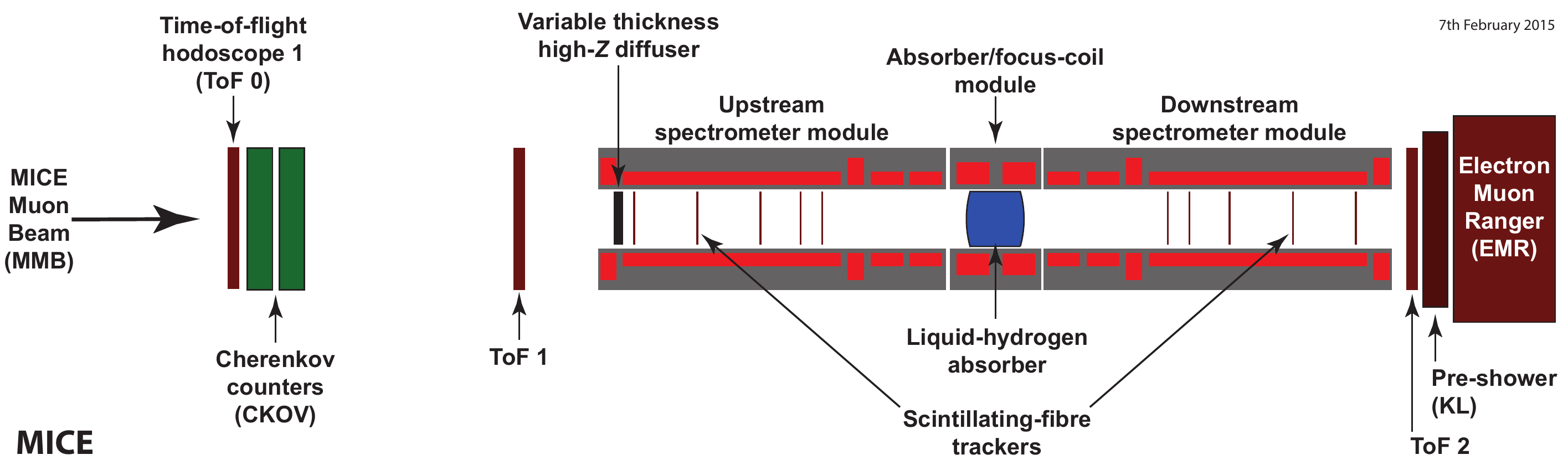}
   \caption{Layout of the MICE Step IV configuration showing the absorber module, tracking spectrometers and detectors for particle identification.}
   \label{fig:mice_step4}
\end{figure*}

\section{MICE}

The Muon Ionization Cooling Experiment (MICE)~\cite{ref:miceweb} has taken data with liquid hydrogen (LH$_2$) and lithium hydride (LiH) absorbers under various optical configurations to make detailed measurements of the scattering, energy loss~\cite{ref:ipac18-tm, ref:ipac18-pf, ref:ipac18-vb} and phase space evolution. A schematic drawing of MICE Step IV is shown in Fig.~\ref{fig:mice_step4}.
An absorber is placed within a superconducting focus-coil module sandwiched between two superconducting spectrometer solenoids instrumented with scintillating-fiber trackers.  Each spectrometer solenoid generates a uniform field in the region of the trackers, and  has a section to match the beam with the adjoining focus coil module. The trackers  provide precise measurements of the emittance upstream and downstream of the absorber. Particle identification upstream of the absorber is performed using two time-of-flight hodoscope (TOF) stations, and two threshold Cherenkov counters. Downstream, contamination from decay electrons is rejected using a TOF, a pre-shower calorimeter and a fully active scintillator calorimeter~\cite{ref:mice-tracker, ref:mice-step1, ref:mice-pions, ref:mice-emr}.  

In this paper the evolution of phase space density is reported for a single configuration (denoted as ``6-140'') of the cooling apparatus where the muon sample has a nominal emittance of 6 mm and momentum around 140 MeV/$c$ in the upstream spectrometer solenoid. The absorber was a 65 mm thick LiH disk.

Since each muon is measured individually, it is possible to select a particle ensemble from the collection of measured tracks. In this analysis, muons have been selected with: 1) longitudinal momentum in the range 135 to 145 MeV/c; 2) time-of-flight between TOF0 and TOF1 consistent with muons in this momentum range; and 3) a single, good quality track formed in the upstream tracker.

\section{PHASE SPACE DENSITY EVOLUTION}

\subsection{Emittance}

The transverse normalized RMS emittance is the most common cooling figure of merit. In a fully transmitted beam, emittance reduction is a clean
signature of the contraction of transverse phase space volume. However, for non-gaussian beams, e.g. partially scraped beams, the RMS emittance is not an accurate measure of the beam phase space. An alternative to RMS emittance is to study the evolution of the density of the ensemble, as it allows for the selection of a well-defined and identical fraction of phase space upstream and downstream of the absorber.

\subsection{Amplitude}

\begin{figure}[!htb]
   \centering
   \includegraphics*[width=\columnwidth]{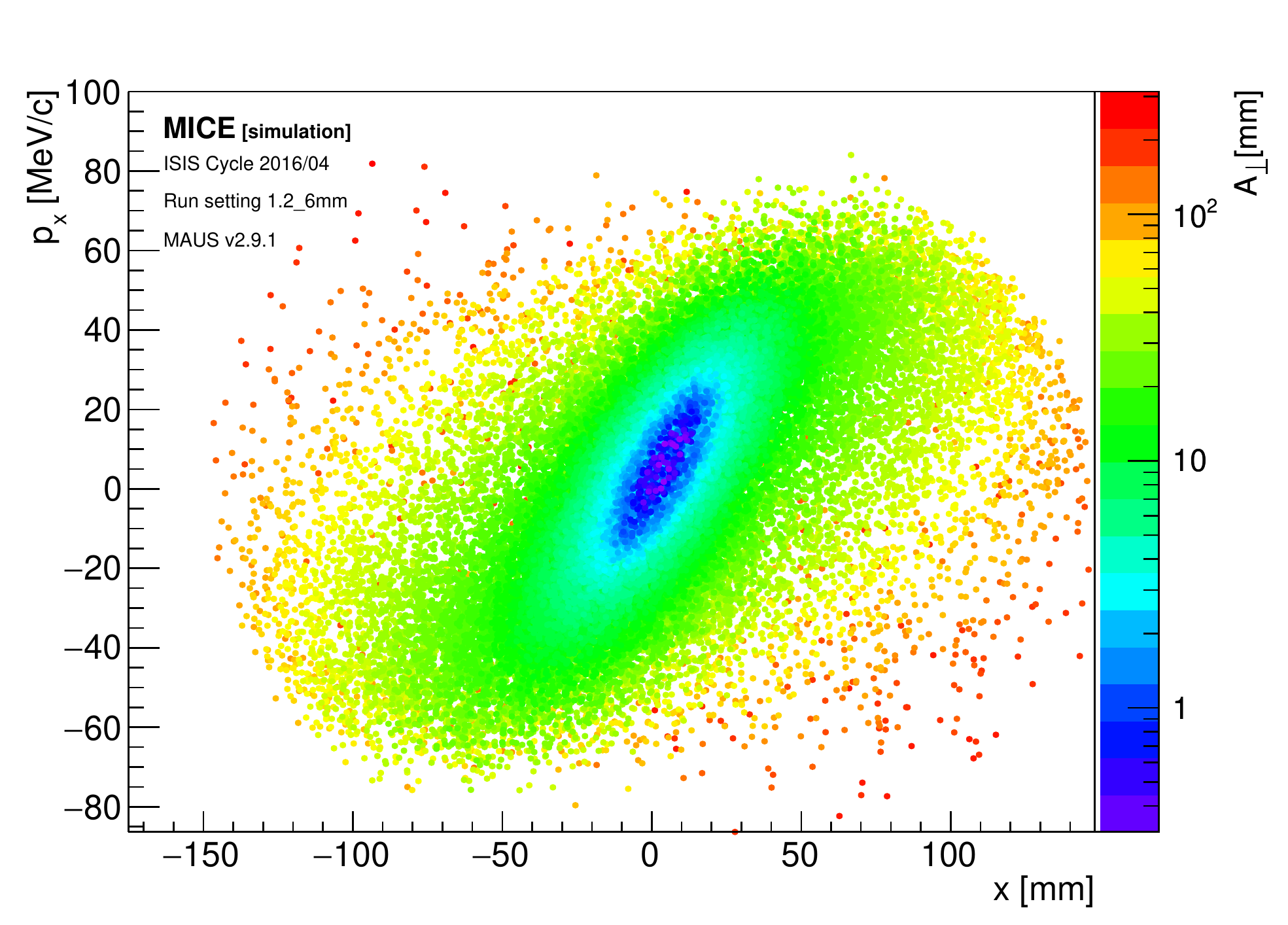}
   \caption{Scatter plot of the particles in the tracker station that is furthest downstream in the $(x, p_x)$ projection. The color scale represents the individual particle amplitudes.}
   \label{fig:amplitude}
\end{figure}

The 4D amplitude of a particle with phase space vector $\mathbf{v} = (x, p_x, y, p_y)$ is given by, 
\begin{equation}
A_\perp = \varepsilon_N \mathbf{(v-\textbf{\si{\micro}})^T}\Sigma^{-1}\mathbf{(v-\textbf{\si{\micro}})},
\end{equation}
where $\textbf{\si{\micro}} = (\left<x\right>, \left<px\right>, \left<y\right>, \left<py\right>)$ is the beam centroid. In order to prevent the tails of the distribution from skewing the core, only those events with amplitude less than $A_\perp$ have been included in the calculation of \textbf{\si{\micro}} and $\Sigma$ for a given event. Thus, high amplitude particles are iteratively removed from the sample. The distribution of muons is represented in Fig.~\ref{fig:amplitude} in the tracker station that is furthest downstream in the ($x,p_x$) projection.
The distribution exhibits a clear high-density Gaussian core of low amplitudes, while the tails are easily identified as high amplitude points.

\subsection{Subemittance}

The $\alpha$-subemittance, $e_\alpha$, is defined as the RMS emittance of the core fraction $\alpha$ of the parent beam. 
For a truncated 4D Gaussian beam of covariance $S$, it satisfies

\begin{equation}
\frac{e_\alpha}{\varepsilon_N} = \frac{|S|^{\frac{1}{4}}}{|\Sigma|^{\frac{1}{4}}} = \frac{1}{2\alpha} \gamma \Big(3,Q_{\chi_4^2}(\alpha)/2\Big),
\end{equation}
with $\gamma(\cdot,\cdot)$ the lower incomplete gamma function and $Q_{\chi_4^2}(\cdot)$ the 4-degrees-of-freedom $\chi^2$ distribution quantiles.

\begin{figure}[!htb]
   \centering
   \includegraphics*[width=\columnwidth]{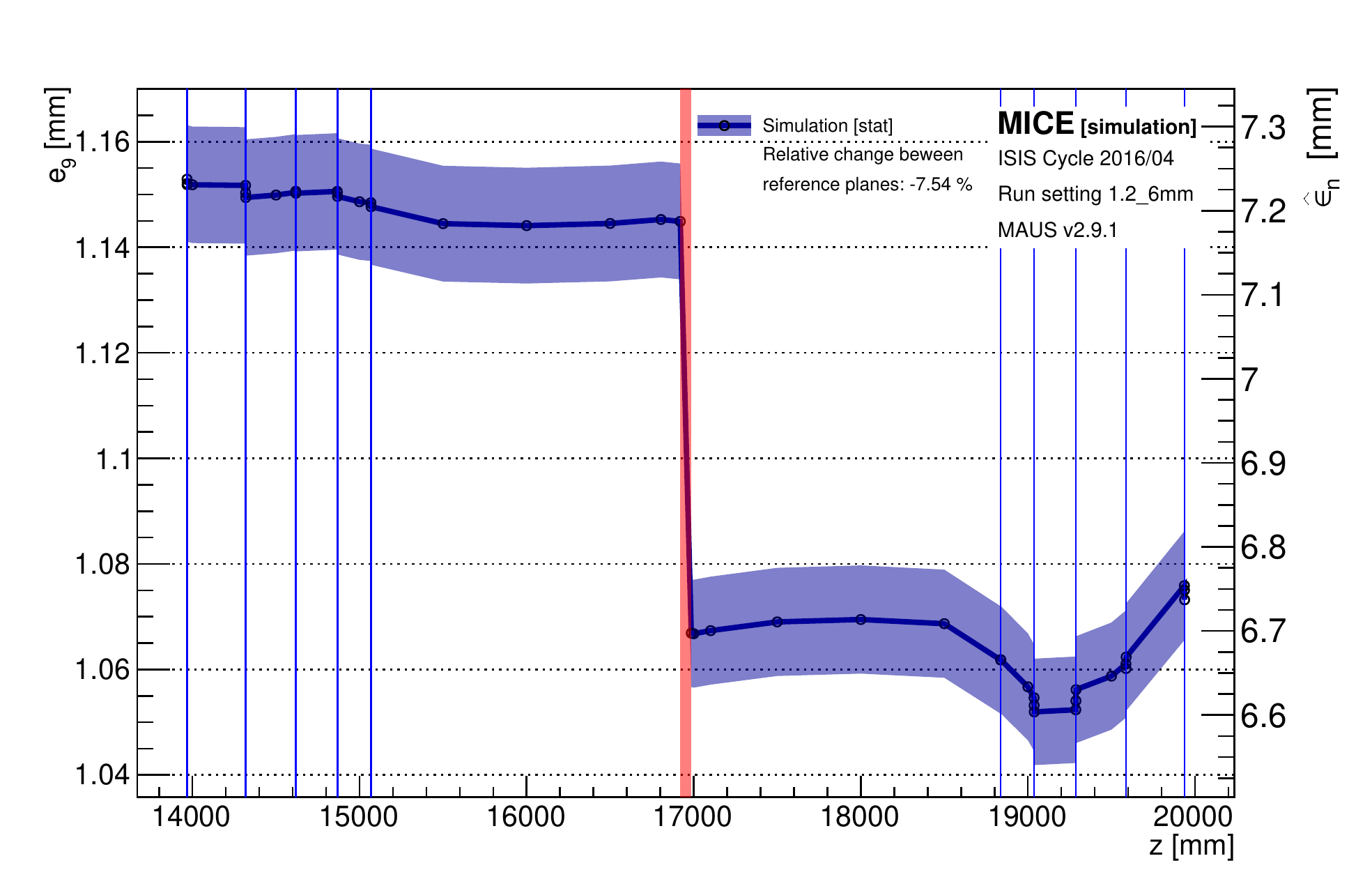}
   \caption{9 \%-subemittance evolution through the the MICE cooling channel.}
   \label{fig:subemittance}
\end{figure}

If an identical fraction $\alpha$ of the input beam is selected upstream and downstream, the measured subemittance change is identical to the
normalized RMS emittance change. The evolution of the 9 \%-subemittance is represented in Fig.~\ref{fig:subemittance}. The choice of
9 \% is natural in four dimensions as it represents the fraction contained within the 4D RMS ellipsoid of a 4-variate Gaussian. This quantity exhibits a clean cooling signal across the
absorber that is unaltered by transmission losses and non-linearities. The only trade-off is that the relative statistical error on $\alpha$-subemittance grows as $\alpha^{-\frac{1}{2}}$ . 
The estimated relative emittance change with this technique is $-$7.54 $\pm$ 1.25 \%, compatible with predictions.

\subsection{Fractional emittance}

The $\alpha$-fractional emittance, $\varepsilon_\alpha$, is defined as the phase space volume occupied by the core fraction $\alpha$ of the parent beam. For a truncated 4D Gaussian beam, $
\varepsilon_\alpha = \frac{1}{2}m^2\pi^2{\varepsilon_N}^2Q_{\chi^2_4}^2(\alpha)$. This volume scales as function of $\alpha$ only and is proportional
to the square of the normalized emittance. For a relative
emittance change $\delta = \Delta\varepsilon / \varepsilon_{N}^{up}$, one yields
\begin{equation}
\frac{\Delta\varepsilon_\alpha}{\varepsilon_{\alpha}^{up}} = \delta(2+\delta) \approx 2\frac{\Delta\varepsilon_{N}}{\varepsilon_N^{up}}.
\end{equation}

\begin{figure}[!htb]
   \centering
   \includegraphics*[width=\columnwidth]{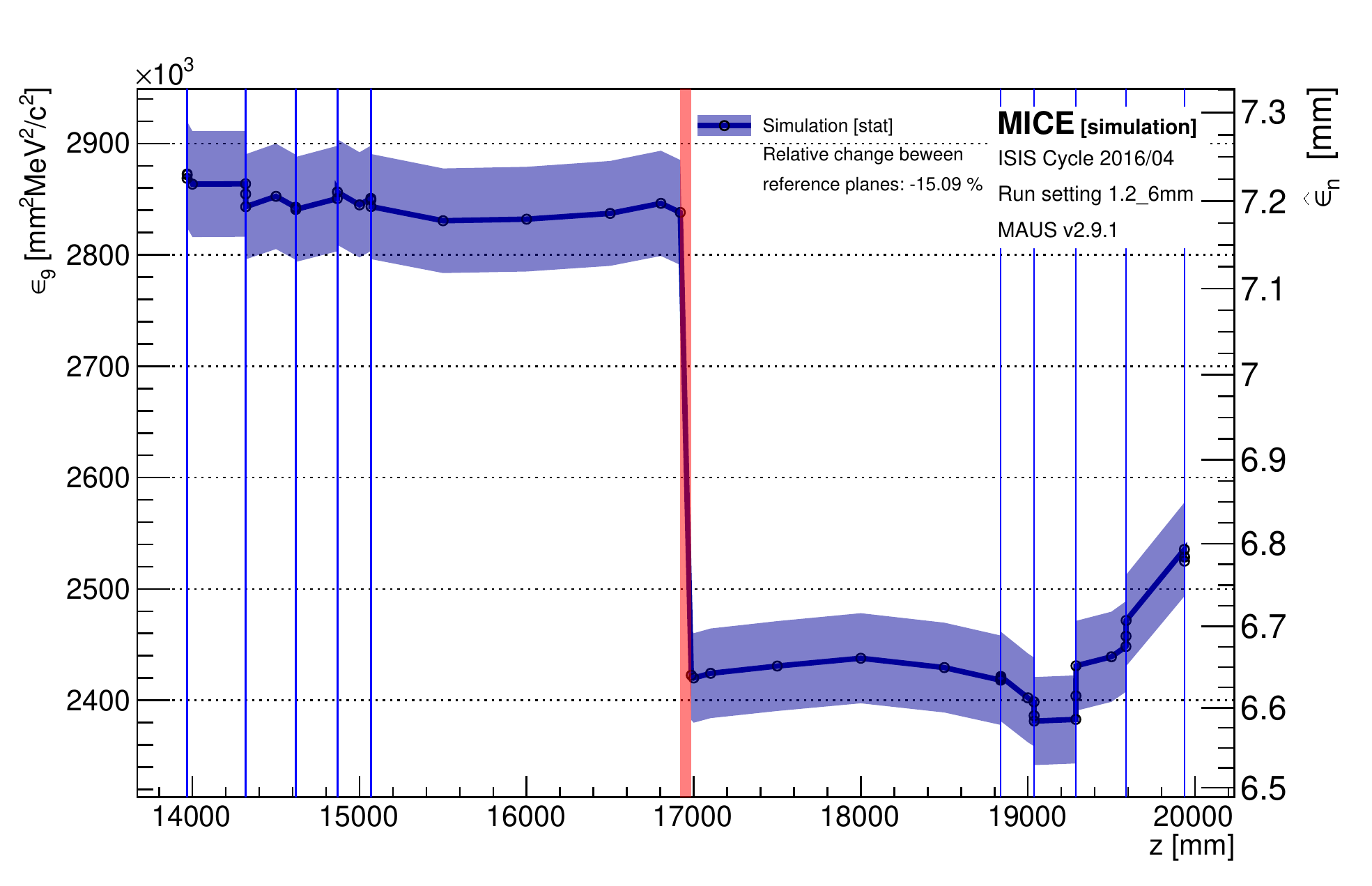}
   \caption{9 \%-fractional emittance evolution through the the MICE cooling channel.}
   \label{fig:fractional-emittance}
\end{figure}

The last approximation holds for small fractional changes. The volume of a fraction $\alpha$ of the beam is reconstructed by taking the convex hull of the selected ensemble~\cite{ref:hull}. Fig.~\ref{fig:fractional-emittance} shows the evolution of the 9 \%-fractional emittance. The estimated relative emittance change with this technique is -7.85 $\pm$ 1.08 \%.

\subsection{Non-parametric Density Estimation}

Unlike parametric density estimation techniques, such as those based on amplitude, non-parametric methods make no assumptions about the probability distributions of the variables being assessed. Among the many classes of estimators that have been developed, three have been considered
in this study: optimally binned histograms, $k$-Nearest Neighbors ($k$NN) and Tessellation Density Estimators (TDEs)~\cite{ref:hogg, ref:knn, ref:nonpar, ref:tde}.
Systematic studies showed that the $k$NN method is the most efficient and robust technique in four dimensions.
For a given phase space vector $\mathbf{v} = (x,p_x,y,p_y)$, we find the $k$ nearest points in the input cloud, calculate the distance, $R_k$, to
the $k^{th}$ nearest neighbor, and evaluate the density as
\begin{equation}
\rho(\mathbf{v}) = \frac{k}{V_k} = \frac{k\Gamma(\frac{d}{2}+1)}{\pi^{\frac{d}{2}}R_d^k},
\end{equation}
with $d$ the dimension of the space, $V_k$ the volume of the $d$-ball of radius $R_k$ and $\Gamma(\cdot)$ the gamma function. 

\begin{figure}[!htb]
   \centering
   \includegraphics*[width=\columnwidth]{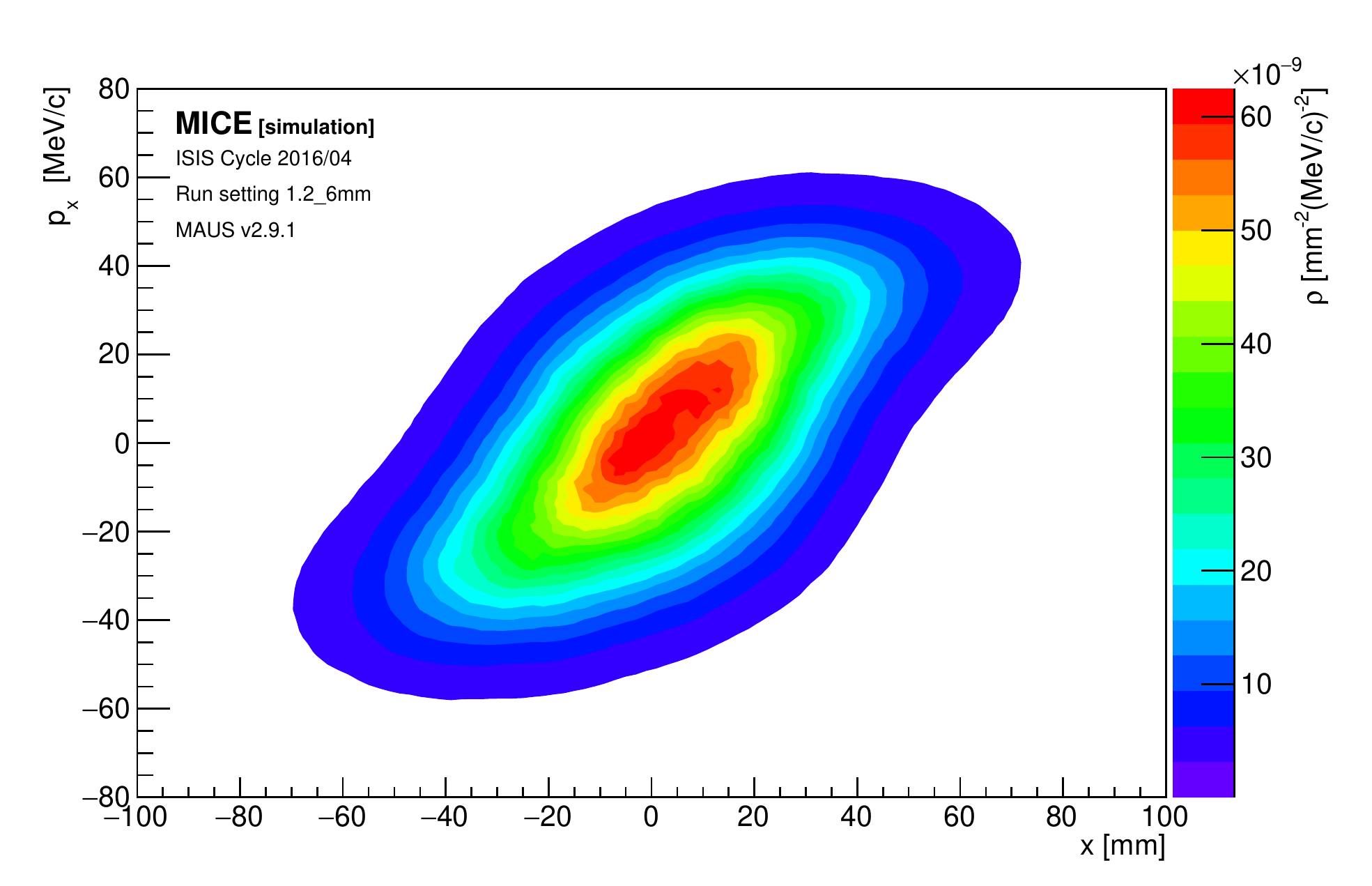}
   \caption{k-Nearest Neighbors estimate of the phase space density in the $(x,p_x)$ projection for $(y, p_y )$ = (0, 0) in the tracker station that is furthest downstream.}
   \label{fig:knn-estimator}
\end{figure}

The choice of parameter $k = \sqrt{N}$ has been shown to be quasi-optimal in general~\cite{ref:nonpar2}.
This estimator is applied to the sample in the tracker station that is furthest downstream and is represented in the $(x,p_x)$ projection for $(y, p_y)$ = (0, 0) in Fig.~\ref{fig:knn-estimator}.
 
\begin{figure}[!htb]
   \centering
   \includegraphics*[width=\columnwidth]{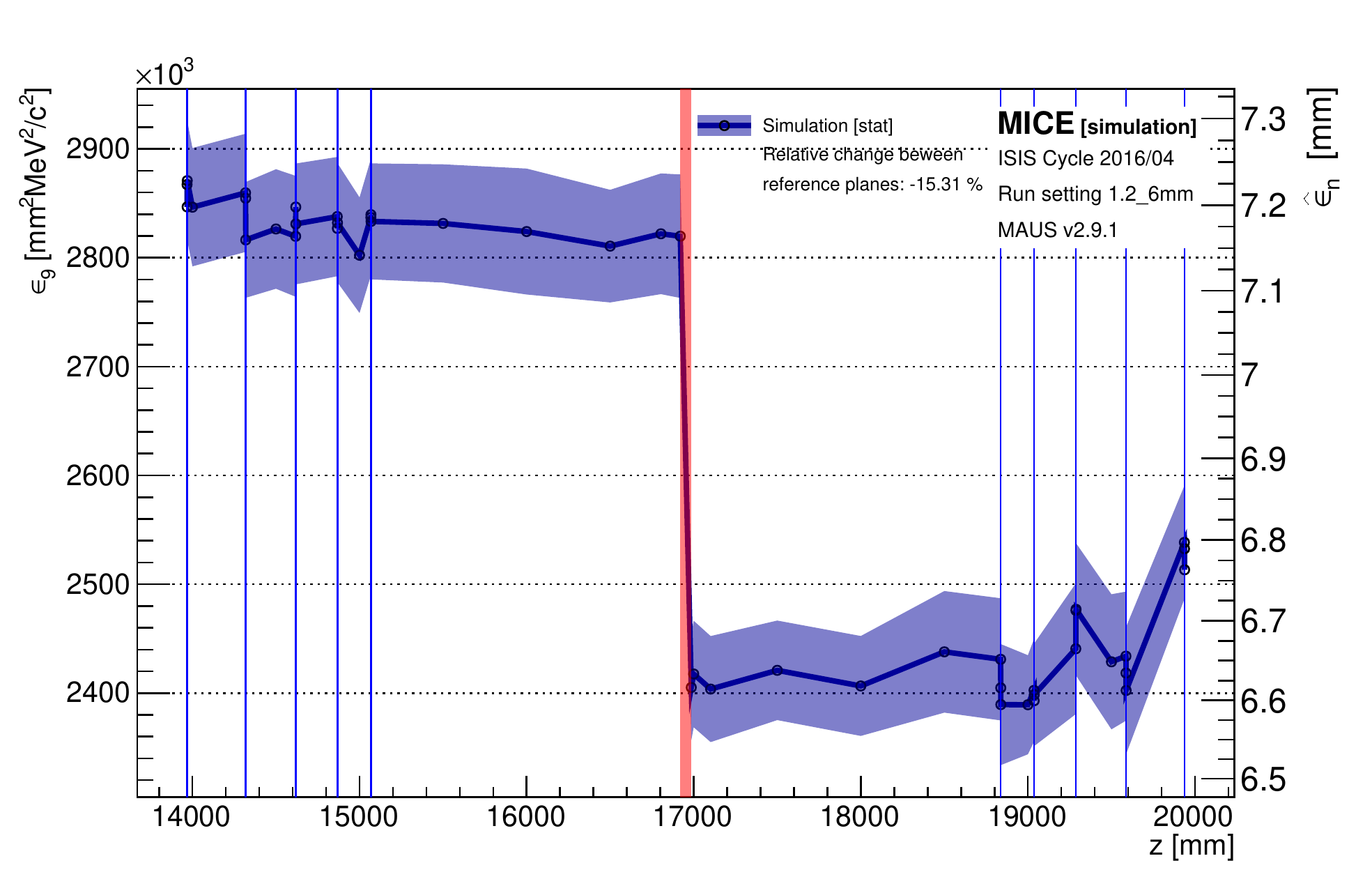}
   \caption{Evolution of the volume of the 9\%- contour of the $k$NN estimate through the MICE cooling channel.}
   \label{fig:knn-evolution}
\end{figure}

This method removes any underlying assumption about a Gaussian core and allows to reconstruct generalized probability contours. The volume of the $\alpha$-contour is the $\alpha$-fractional emittance, as defined above. An MC method is used to reconstruct the volume of a contour: we select the densest fraction $\alpha$ of the input points and record the level of the lowest point, $\rho_\alpha$. N random points are sampled inside a box that encompasses the contour and we record the amount $N_\alpha$ with a density above $\rho_\alpha$. The volume of the contour is simply $\varepsilon_\alpha = N_{\alpha}V_{box}/N$, where $V_{box}$ is the volume of the 4d-box. The 9 \%-contour volume evolution is represented in Fig.~\ref{fig:knn-evolution}. The estimated relative emittance change with this technique is -7.97 $\pm$ 1.63 \%.

\section{CONCLUSION}

While the traditional normalized RMS emittance measurement is vulnerable to transmission losses and non-linearities in the particle ensemble, density estimation techniques provide the most viable option to recover quantitative cooling measurements. Amplitude-based techniques -- subemittance and fractional emittance -- rely on a well known quantity to select and study an identical fraction of the beam upstream and downstream of the absorber. Nonparametric density estimators allow to go one step further in removing any assumption on the underlying distribution. Both approaches yield compelling results when applied to a poorly transmitted and highly non-linear beams in a realistic simulation of one of the MICE cooling channel settings.
\ifboolexpr{bool{jacowbiblatex}}%
	{\printbibliography}%
	{%
	

} 

\end{document}